# Limited-angle x-ray nano-tomography with machine-learning enabled iterative reconstruction engine

Chonghang Zhao, Mingyuan Ge, Xiaogang Yang, Yong S. Chu, Hanfei Yan*

National Synchrotron Light Source II, Brookhaven National Laboratory, Upton, NY 11973, USA

*email: hyan@bnl.gov

Zhao, Chonghang:  chonghang.zhao@gmail.com
Ge, Mingyuan: mingyuan@bnl.gov
Yang, Xiaogang: yangxg@bnl.gov
Chu, Yong: ychu@bnl.gov
Correspond author:
Yan, Hanfei
Brookhaven National Laboratory
National Synchrotron Light Source II
Bldg 743/Rm 103
Upton, NY 11973
Tel: 631-344-7097
hyan@bnl.gov

## ABSTRACT

A long-standing challenge in tomography is the 'missing wedge' problem, which arises when the acquisition of projection images within a certain angular range is restricted due to geometrical constraints. This incomplete dataset results in significant artifacts and poor resolution in the reconstructed image. To tackle this challenge, we propose an approach dubbed Perception Fused Iterative Tomography Reconstruction Engine, which integrates a convolutional neural network (CNN) with perceptional knowledge as a smart regularizer into an iterative solving engine. We employ the Alternating Direction Method of Multipliers to optimize the solution in both physics and image domains, thereby achieving a physically coherent and visually enhanced result. We demonstrate the effectiveness of the proposed approach using various experimental datasets obtained with different x-ray microscopy techniques. All show significantly improved reconstruction even with a missing wedge of over 100 degrees—a scenario where conventional methods fail. Notably, it also improves the reconstruction in case of sparse projections, despite the network not being specifically trained for that. This demonstrates the robustness and generality of our method of addressing commonly occurring challenges in 3D x-ray imaging applications for real-world problems.

## 1. INTRODUCTION

X-ray tomography is a ubiquitous imaging tool for nondestructive characterization of an object's interior in three dimensions (3D), with a broad range of applications in science and technology[1-5]. This technique entails acquiring a series of projection images as the object rotates, followed by the application of a computer algorithm for volume reconstruction. The reconstruction quality depends on factors such as the number of projections, angular coverage, and the chosen algorithm[6]. It is also affected by experimental imperfections such as noise and misalignment error. The widely acknowledged Crowther criterion establishes the required number of projections over a complete circle for achieving high-quality reconstruction[7]. However, certain scenarios impose significant limitations on the projection angle due to geometric constraints, such as the imaging of integrated circuits (ICs) or using *in situ* cells with limited angular access. In particular, the planar geometry of ICs substantially restricts the maximum rotation angle due to rapid attenuation increase, resulting in the 'missing wedge' problem in tomography[8]. This lack of angular coverage leads to severe artifacts and distortion in the reconstruction.

Mathematical methods can help alleviate the impact of data imperfection. Conventional tomography reconstruction algorithms fall under two categories: direct methods based on either back propagation or Fourier central slice theorem[9], which offer speed but are sensitive to data imperfection, and model-based iterative methods, which are more computationally intensive but less prone to artifacts[10-12]. While model-based iterative methods exhibit capability in handling data with a small missing angular range, they struggle when the missing wedge is substantial (e.g., over 40 degrees). Conventional methods for limited-angle tomography have their limits in dealing with missing information. Recently, the emergence of deep neural networks (DNNs) has presented a promising solution to this challenge, demonstrating success in applications previously deemed challenging or impossible[13-20].

DNN-assisted reconstruction methods offer various pathways, with post-processing being the most straightforward. In this case, DNNs are trained to remove artifacts and enhance image quality after the initial reconstruction[13,14,16]. However, their correcting power is constrained without a feedback mechanism. DNNs can also be trained to achieve sinogram inpainting by synthesizing missing data not being measured during acquisition. While demonstrated in sparse-view[21] and limited-angle tomography[19], the reliance on synthetic data introduces uncertainties about reconstruction fidelity. Efforts to address this challenge involve dual-domain corrections, using two DNNs applied in sinogram and image domains, yet challenges persist[20,22].

An alternative approach involves DNNs replacing the inverse radon transform, directly mapping information from the detector domain to the image domain. In this approach, DNN needs to learn the underlying physics and often requires a large amount of training data[15]. To alleviate the need of training data, Yang *et al.* introduced a Generative Adversarial Network (GAN) with a self-training approach, leveraging a physical model as feedback scoring mechanism[18]. Barutcu *et al.* further improved the GAN-based self-training approach by incorporating TV-based regularization, demonstrating enhanced image quality and convergence properties[23]. While these DNN-based inverse transforms outperform conventional solvers in specific cases, questions arise regarding their generalization ability.

A final category comprises model-based deep learning, aiming to combine the accuracy, generality, and stability of physical models with the capacity of DNNs to capture implicit and complex knowledge. One can either embed a physical model into a network by unrolling the iterative algorithm into feed-forward layers and optimize the network in a data-driven manner or integrate a DNN into an iterative process as a regularizer. The former demonstrated high convergence efficiency owing to learnable tuning parameters in an iterative solving process[24]. However, it requires pre-defined iteration steps, therefore less flexible in dealing with different scenarios. The latter approach uses a DNN as a regularizer in an iterative algorithm[25], enforcing the solution with desired properties. One can encode perceptual knowledge of samples in the neural network from the training but can be computationally intensive.

The choice of DNN usage in limited-angle tomography depends on application-specific considerations such as computation speed, model complexity, training requirements, and generality. In this work, we developed a machine-learning-aided reconstruction method emphasizing robustness, generalization, and fidelity—properties essential for scientific applications. We adopt the model-based deep learning approach where DNN is used as a regularizer, motivated by the concise and explicit nature of physical models, and the significance of data consistency and physical constraints in scientific problems.

Our implementation comprises two key components: a U-net encoder-decoder convolutional neural network[26], and Alternating Direction Method of Multiplier (ADMM) as the iterative engine that integrates the physical solver and the network[27]. The U-net effectively captures features across multiple length scales, and we modified the latent space to enhance its receptive field for better performance. Through training on a synthetic dataset consisting of various objects, the modified U-net acquired perceptual knowledge about both characteristic features of samples and artifacts/distortion associated with limited-angle tomography. ADMM splits the original optimization problem into two sub-problems, one in physics and the other in image domains. A plug-and-play scheme is taken for the image sub-problem, so that the regularizer function does not need to be defined explicitly. In each iteration, a corrupted reconstruction is improved by the neural network, and then feedback to the physical solver as the initial guess of the solution. The interplay between the physics and image domains ensures a coherent integration of the perceptual knowledge into the physics-aware solution and prevents introducing additional artifacts from the neural network itself. Our contribution is twofold: (1) the design and training of a DNN capable of effectively correcting corrupted reconstructions while ensuring stability, and (2) its integration into a conventional iterative solving engine as a "smart" regularizer.

Dubbed Perception Fused Iterative Tomography Reconstruction Engine (PFITRE), our method's effectiveness is demonstrated across various experimental datasets taken at two different beamlines, including IC samples, battery electrodes, and porous metals, with projections limited to as low as 60°. Benchmarking against previously published datasets also demonstrated its superior performance. Tests for sparse tomography further confirm its generalization ability even though the network was not specifically trained for this scenario. This highlights PFITRE's broad applicability across diverse samples, modalities, and imaging modes without requiring retraining—an achievement not demonstrated by other methods.

2. RESULTS

In this study, our objective is to tackle the missing-wedge challenge in tomography. As depicted in Figure 1(a), the impact of missing measurements in certain angular range can be understood in reciprocal space where a section is unsampled. This leads to a well-known 'missing wedge' problem in tomography. For x-ray nano-tomography, it can be measured either in scanning or full-field mode, as shown in Figure 1(b). Here we introduce an iterative reconstruction method incorporating machine learning-based regularization for limited-angle tomography. The schematic image of our method is shown in Figure 1(c). Our work is developed based on the Alternating Direction Method of Multipliers (ADMM)[27]. By projecting optimization into two domains and iteratively updating solutions, we aim to obtain a solution that satisfies constraints in both domains. In one domain, the gmres linear solver in SciPy[28] was used to ensure a constraint of physical accuracy. Here, the radon transforms of the reconstructed object, denoted by x, must closely match the measured sinogram. A well-informed initial guess, "a warm start", can enhance reconstruction performance. In the other domain, the constraint was applied by CNN-based regularization. We use variable z to represent the object in this domain. Detailed ADMM implementation can be found in the methods section.

In the following paragraphs, we first discuss the performance of various networks on synthetic datasets, applying them as post-processing methods in image domain. After identifying the optimal network architecture, we integrate it into the ADMM framework, developing PFITRE for effective artifact removal and image correction with projection data.

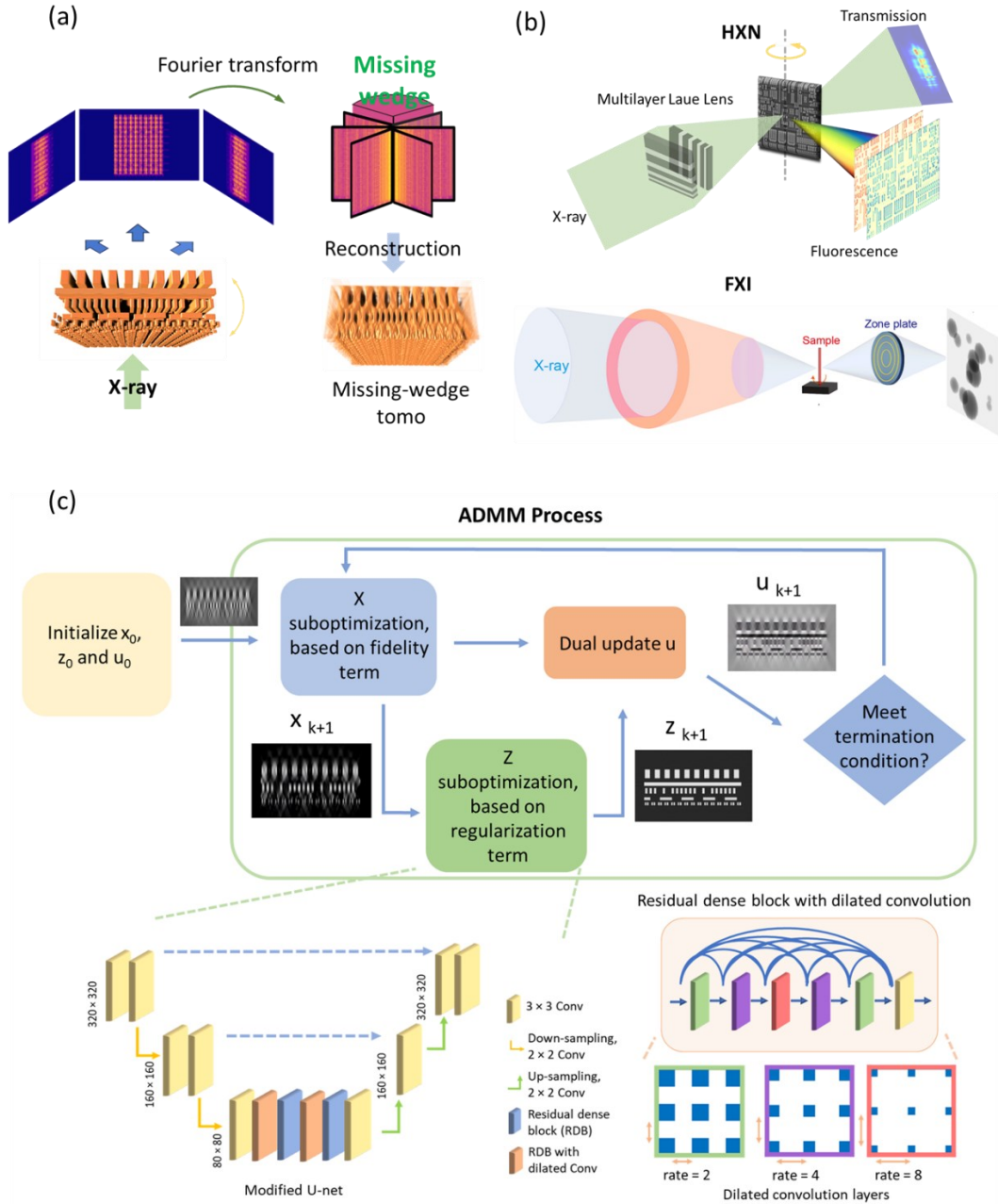

Figure 1. Schematics illustrating the "missing-wedge" problem, data acquisition mode, and the use of an ADMM-based iterative method that incorporates a modified U-net to correct, inpaint and restore a corrupted tomography reconstruction in an iterative manner. (a) Illustrates the 'missing wedge' problem encountered in tomography when the projections are not over 180°. A section of the reciprocal space with a wedged shape is unsampled due to the missing measurements in certain angular range. (b) Depicts the experimental setup for scanning fluorescence tomography at hard x-ray nanoprobe beamline (HXN) and full-field absorption tomography at full-field imaging beamline (FXI) of National Synchrotron Light Source II (NSLS-II). (c) Shows the flowchart of PFITRE that incorporates a DNN into ADMM algorithm. The DNN has a modified U-net architecture, featuring the newly introduced Residual in Residual Dense Block with dilated convolution.

## 2.1. Evaluation of network performance on synthetic test data

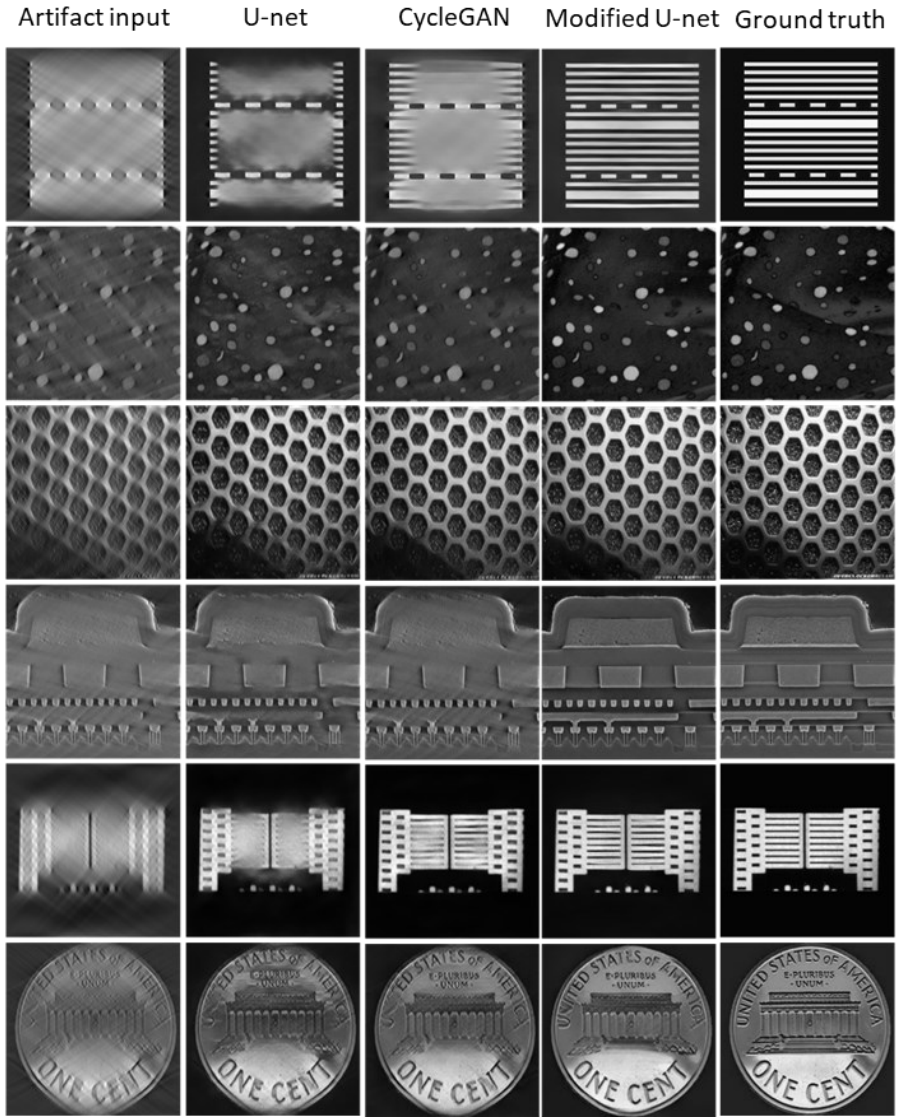

Figure 2. Images contain artefacts resulted from limited-angle tomography (first column), their corrections using different network architectures (2nd-4th column), and ground truth (last column).

The comparison of network performance in image domain between our modified U-net with other conventional U-net-based architecture on the synthetic test dataset is presented in Figure *2*. All the networks were trained with the same training dataset and followed the same procedure. The synthetic images from various categories, after being reconstructed by a linear solver are displayed in the first column. Missing-angle-induced artifacts are particularly noticeable in the category of integrated circuit images, where many of the line-shaped features appear broken and distorted. The image quality in this category is inferior to that of other categories, and correcting artifacts in these images is a challenging task. This can be understood by considering that line features are concentrated within a very limited angular range in reciprocal space. The absence of projections within this specific angular range results in the loss of critical information, leading to severe distortions in the reconstructed images. In contrast, features in natural images are distributed across a wider angular range in reciprocal space. Some information is preserved even with limited projections, resulting in better results when reconstructed using a linear solver.

Corrections made by different networks are displayed in the 2$^{nd}$ to 4$^{th}$ columns of Figure *2*. The conventional U-net and CycleGAN architecture were able to partially correct artifacts in the pattern and natural images. However, they failed to correct broken line artifacts, and could not recover the boundaries of objects in the images. This limitation is attributed to their architecture, which is based on 3×3 convolutions with a limited receptive field. By implementing RRDBs with dilated convolution layers, our network can expand the receptive field and recover information at different length scales in the images. This design enables the successful recovery of IC images in cases where the central region is missing and blurry. It produces results that closely resemble the ground truth after one correction.

Table 1. Qualitative assessment of network architectures by different metrics on test dataset. Our modified U-net outperforms the original U-net architecture and CycleGAN in four different metrics.

|  | $L_1$ × 10$^{-7}$ | VGG × 10$^{-7}$ | SSIM | PSNR |
| --- | --- | --- | --- | --- |
| U-net | 0.998 | 5.386 | 0.647 | 24.496 |
| CycleGAN | 0.753 | 4.289 | 0.711 | 26.433 |
| Modified U-net | 0.668 | 3.746 | 0.740 | 27.539 |

In Table 1, four quantitative metrics including SSIM, PSNR, L1, and VGG, for analyzing the artifact correction by different networks are presented. The analysis was conducted on a test dataset containing over 4000 images reconstructed from a limited angle range of 90° - 160° with a 1° angular step, which have never been seen by the network. The quantitative measures of the correction align well with observations, and our modified U-net outperforms other U-net-based networks.

Due to the way of how DNN is applied in the iterative engine, one important property of the network is stability. We enforce that by introducing GT-GT pairs in the training dataset and the identify loss. Their effects are shown in Figure S1. We plot outputs from the 1$^{st}$, 5$^{th}$, and 10$^{th}$ iterations in Figure S1 (a) with network trained under various conditions. Without GT-GT pairs or identity loss, image quality deteriorates after iterations. In contrast, when the network was trained with identity loss or GT-GT pairs in the training dataset, it remains stable. The quantitative analysis of differences between each iteration output, and between the output with the ground truth, are plotted in Figure S1(b) and (c). It can be observed that after the inclusion of GT-GT pairs in the training dataset and the integration of identity loss, the network's output is closer to the ground truth, and its convergence properties improve. This is critical because the network will be incorporated as a regularization term into an iterative engine and applied repetitively. Additional iterative tests on the test dataset are shown in Figure S2.

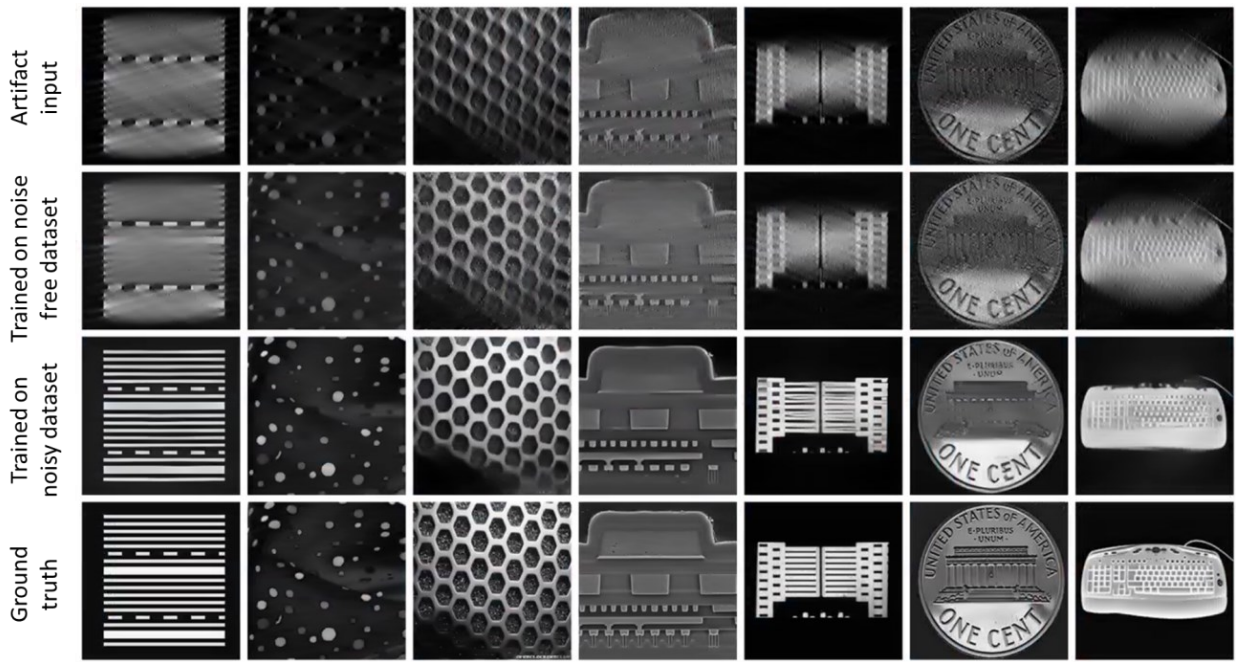

Figure 3. Corrections in image domain on synthetic data with limited-angle artifacts, noise, misalignment, and normalization issues. The 2$^{nd}$ row shows the output from a network trained with the limited-angle artifact only. The 3rd row shows output from a network trained with all mentioned artifacts.

In real-life experiments, besides instrumental noise or background signals, artifacts due to other data imperfection may be present, and neural networks are known to have limited capabilities when dealing with unseen datasets. This makes training networks on synthetic datasets more challenging. To address this issue, we generated a new synthetic dataset that goes beyond missing-angle artifacts and includes additional potential artifacts encountered in tomography measurements, such as noise, misaligned projections, and intensity variations among projections. We conducted transfer learning on this new synthetic dataset using a network pre-trained on a dataset that only considered limited-angle artifacts. In Figure 3, we demonstrate the effect of network performance with and without transfer learning on a new dataset. It is evident that while the network trained on a noise-free dataset exhibits the ability to correct artifacts induced by missing angles in noisy images, its performance is suboptimal, particularly in the case of IC images. In contrast, a clear improvement is observed in models trained on noisy datasets, which yield results close to the ground truth. Therefore, simulating an environment that closely resembles real-life conditions and employing transfer learning on noisy datasets are critical. In the subsequent testing of experimental results, we utilized the network trained on the noisy dataset.

2.2. Limited-angle tomography on experimental data for IC samples

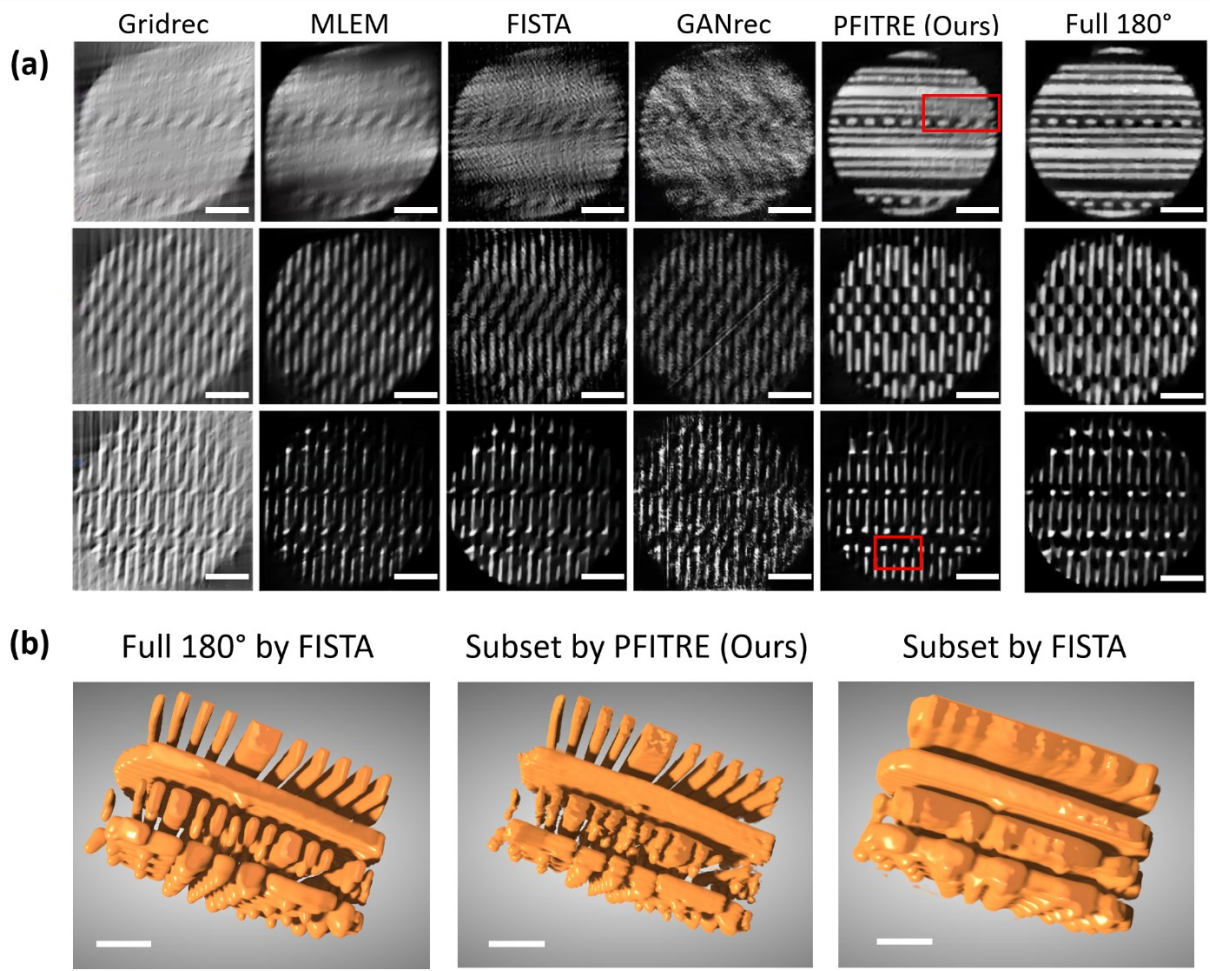

Figure 4. Tomography reconstructions with a subset of projections using different methods. (a) A comparison of reconstructions on three layers with Gridrec, MLEM, FISTA, GANrec method, our method (PFITRE), and the full dataset spanning 180° reconstructed by FISTA. All scale bars indicate 500 nm. It should be noted that although PFITRE produces a visually appealing solution even when the missing-wedge is substantial, there could be still observable differences at some locations from the ground truth, as indicated by red squares. (b) 3D volume rendering with cross-sections along the thickness from the full angular range dataset, the subset reconstructed by our method, and the FISTA method.

While tests on synthetic datasets show excellent correction power of the network even with one-time post-processing, its effectiveness on experimental data is always a challenge because unseen patterns/artifacts may be present. Incorporating the network into an iterative solving engine can significantly extend its generalization ability and improve robustness since all non-physical corrections will be discarded.

The first experimental dataset we tested is Cu XRF tomography of the cylindrical IC sample with projections spanning 180°. The reconstruction from the full dataset was treated as the ground truth for reference. To simulate limited-angle conditions, we selected projections covering 84°, a significant reduction of the angular range, and performed reconstruction using different methods.

In Figure 4, we compared results from PFITRE with those from other algorithms, including Gridrec, Maximum Likelihood Expectation Maximization (MLEM), Fast Iterative Shrinkage/Thresholding Algorithm with total variation regularization (FISTA), and the machine learning-based GANrec method[9,18,29,30]. Results for other elements are shown in Figure S3. Three representative layers in the reconstructed volume showing characteristic patterns of IC samples are shown in Figure 4(a). As expected, iterative reconstruction methods such as MLEM and FISTA outperformed the Fourier transform-based Gridrec. GANrec, although is machine-learning powered, did not perform well in this case, maybe due to the fact that a "warm-start" with pre-trained parameters suitable for recovering IC samples is required. As evident, there are substantial distortions/artifacts in these reconstructions. Particularly for the first layer, because the critical information of these line structures lies in the missing wedge, nothing recognizable was reconstructed.

In contrast, PFITRE yielded significantly better results in all three cases, thanks to the incorporation of prior perceptual knowledge into the iterative engine. Even for the most challenging case (the first layer), the algorithm was able to generate an image preserving shape and structure very close to the ground truth. Interestingly, it even produced shaper images than those from the complete dataset for the second and third layers, indicating its ability to improve reconstruction quality in general. It should be noted that although PFITRE produces a visually appealing solution, there are still observable differences at some locations from the ground truth, as indicated by red squares in Figure 4(a). This is because when the missing wedge is substantial, the data consistency term cannot reject a solution that may be equally probable with the ground truth from a random start. Therefore, a "warm-start" with an initial guess close to the ground true is important when the missing information is significant.

In Figure 4(b), we presented 3D renderings of the reconstructed volume using FISTA and the full dataset, PFITRE and the subset, and FISTA and the subset. Reconstruction from PFITRE successfully resolves metal connectors in the IC sample with a very limited projection range of 84°, which was otherwise impossible using conventional methods. An equally good demonstration on another dataset with the same angular of 84° but a reduced number of projections (every 2°) was provided in Figure S4 and supporting Video. One imperfection seen is the zigzag boundaries for the fine lines. This is mostly because currently reconstruction is done layer by layer, which may lead to an inconsistency of intensity between layers. The implementation of a 3D kernel-based neural network with a corresponding iterative method is expected to provide further improvement, with a price of more computation.

### 2.3. Improvement over iterations

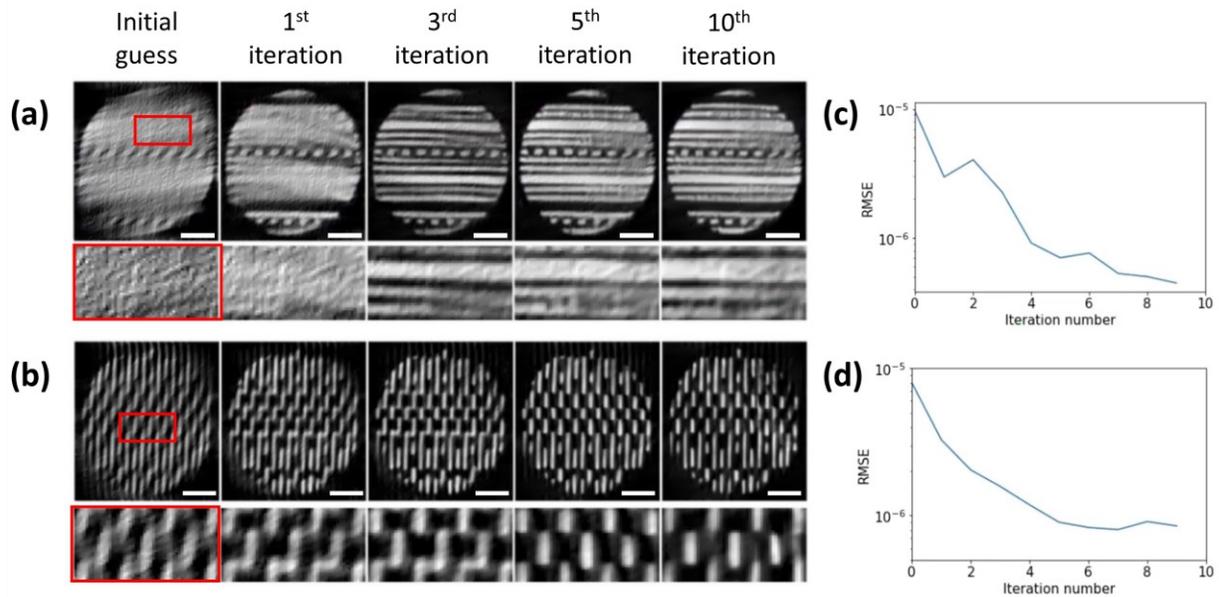

Figure 5. Gradual improvement over iterations on a subset of projections from the experimental results. From left to right: the initial guess, the image corrected after a different number of iterations by our method. The plot shows the Root Mean Square Error (RMSE) that quantifies the relative change of the reconstruction after each iteration. All scale bars indicate 500 nm.

The interplay between the network and the data consistency term ensures that any non-physical changes are removed, and the iterative process gradually improves the solution. Figure 5 illustrates the intermediate results and how they are improved over iterations for the same two layers in Figure 4. Evidently, one-time correction (the 1st iteration) is not sufficient to correct a severally corrupted reconstruction, even though the network was trained to restore it to the ground truth. This could be due to the complexity of artifacts seen in experimental data. It is very difficult to include all possible scenarios in the synthetic training dataset that one can encounter in an experiment. As a result, the correction will not be as good as with that on the verification data. While the improvement of the first iteration is most dramatic, further enhancement leading to shaper edges and finer details are observed with more iterations.

Furthermore, the iterative process was effective in removing artifacts introduced by the network itself. In Figure 5(b), PFITRE was able to remove non-physical horizontal lines that connects vertical ones in the following iterations, making the algorithm less error-prone. The network attempted to restore the input image to what is seen in the training dataset, without knowing whether it is correct or not. The data consistency term provides feedback to verify the correctness and remove any non-physical changes. With iterations, once converged, the solution will meet both the expectation from prior perceptual knowledge and satisfy the physical model.

In Figure 5(c) and (d), the convergence rate measured by the Root Mean Square Error (RMSE) of two consecutive iterations is shown in each case. Note that the value was normalized by the image size. They offer a quantitative assessment of needed iterations. As can be seen, 5 iterations usually are sufficient to reach a convergence.

## 2.4. Impact of the orientation of the missing wedge

When the missing wedge is the same, how it is oriented with respect to the structure can also play a role in determining the reconstruction quality of an IC sample. Here we use a tomography dataset consisting of projections over 128° for a planar IC sample to demonstrate this effect. In doing so, we reduce projections from 128° down to 64°, and orient the missing wedge in two ways, symmetric and asymmetric with respect to the 0° angle, which is along the thickness axis (or the vertical axis of the reconstructed image). In Figure 6 we show reconstructions from projections symmetrically and asymmetrically reduced. The corresponding reciprocal space plots of different projection distribution is shown in Figure S7. Even though PFITRE was able to reconstruct a sharp and clean image in all cases, the asymmetric reduction tends to introduce more artifacts, particularly when the projection range is reduced to 64°. One might think going high angle on one side can help obtain frequency information along the vertical axis in reciprocal, but the asymmetry of the measurement leads to a very skewed initial guess, making the final recovery more difficult. When an object with certain symmetry like an IC sample is under investigation, symmetric projections are desirable, unless an algorithm can produce a skewness-free initial guess. The impact of the initial guess is discussed in more detail in the supplementary materials.

One may notice that the smallest projection range tends to produce the "cleanest" image. This is because with less data, the result is more "guessed" by the network and less affected by the measured data. Tweaking the weight of the regularization term can help achieve the desired balance. In Figure S8 we discussed the impact of the weight.

It is worth noting that reconstruction with projections over 64° is not included in the training dataset. This indicates the method's robustness beyond the training range.

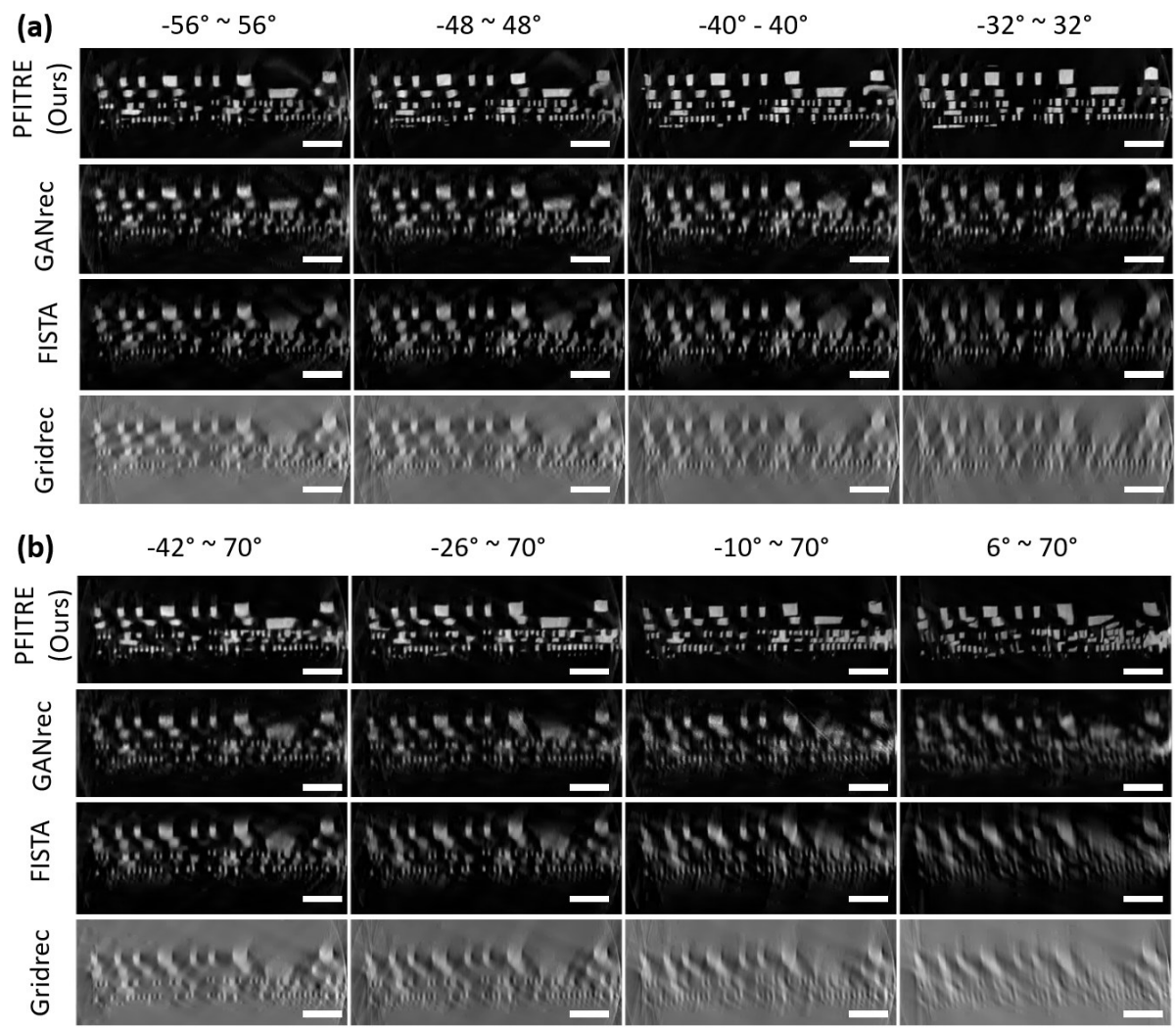

Figure 6. Comparison of reconstructed planar IC samples using different methods with varying angular coverage. The reconstruction methods employed include PFITRE (ours), GANrec, FISTA, and the Gridrec method. (a) Reconstruction from projections is symmetrically distributed around 0° at different angular ranges, with 0° defined as the planar sample is perpendicular with the incident beam. (b) Reconstruction from asymmetrically distributed projections, covering angles ranging from 64° to 112°. The collected angle ranges from -58° to 70°, covering 128° in total. All scale bars represent a length of 500 nm.

2.5. Generalization to other samples and sparse datasets

For scientific measurements, we often encounter unseen samples or experimental conditions that are not included in the training dataset. For broad scientific applications, it is important to validate the generality and robustness of the method. We tested on a tomography dataset for $LiMn_2O_4$ electrode in Li-ion batteries acquired at the Full-field X-ray beamline (18-ID, NSLS-II). Neither the sample, nor the modality, was included in the training dataset. Comparing results are shown in Figure 7. PFITRE again shows significantly improved reconstruction quality than others when projection angles are limited.

Particularly when the range is reduced to 70°, although there are some loss or subtle changes of fine details, it still produces a sharp image with overall high fidelity. As discussed previously, when the missing information is too substantial, there is a chance that a recovered feature is not real because the fidelity term cannot discern it from the ground truth with limited data. They may be equally probable based on the likelihood function. Therefore, care must be taken when interpreting fine features in such an extreme case. Nevertheless, overall features of the sample are reconstructed correctly. In comparison, FISTA yields image with a strong distortion and blurring. GANrec partially corrects the distortion, but it struggles to remove artifacts and recover fine details. Because it directly maps the sinogram to the object, the network may not capture the perceptual knowledge of the sample and use it to enhance the image quality.

To provide a fair benchmark against GANrec, we applied PFITRE to the same dataset published in reference 18, with the results presented in Figure S9. The superior performance of PFITRE is evident. Additionally, we conducted another test using data from the Helsinki Tomography Challenge 2022[31], where these objects had not been previously seen by the network. High-quality reconstructions were achieved when the projections exceeded 75° (Figure S10).

We also tested the impact of sparsity on this dataset by reducing the number of projections. As shown in Figure 7(b), there is no apparent degradation when the number of projections is halved, and the impact is more pronounced when the reduction is greater than 4. There is a bit of loss of resolution and clarity in the case of projection over 180° at every 8°, where only 24 projections are used. Considering the big gain (8x) in time, this might well be acceptable. More details are lost, and blurring is more apparent as the angular range is reduced, as expected. In general, it requires about 20 projections, i.e., 4° step over 70°, 6° step over 120°, and 8° step over 180°, respectively, to preserve the correct shape of the features. Considering that the network is not trained with sparse datasets, we believe this indicates PFITRE's generalization ability and potential to tackle sparse problems as well. By expanding the training dataset, we expect a further improvement in its performance. In Figure S12 of the supplementary material, another example for porous Cu sample is presented.  Similar conclusions can be drawn.

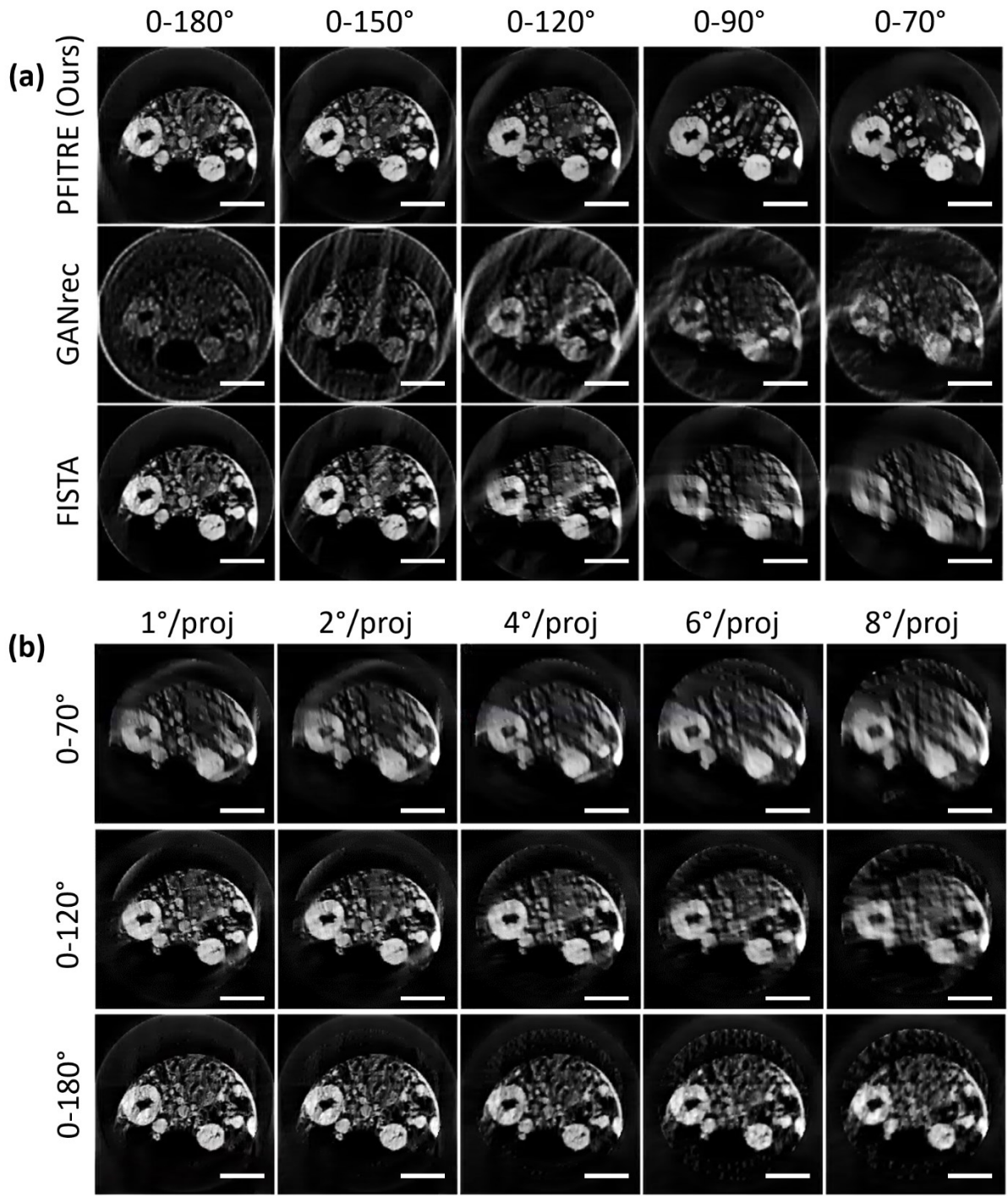

Figure 7. Reconstruction of the LiMn$_2$O$_4$ electrode in a Li-ion battery using transmission X-ray microscopy with projections covering various angles. The reconstruction methods used include our iterative method, GANrec, and FISTA. (a) A comparison of reconstructions with projections covering an angular range from 70° to 180° using different methods. (b) Evaluation of performance on sparse datasets with varying angular coverage and sparsity levels. All scale bars represent a length of 5 μm.

3. DISCUSSION

We have developed PFITRE, an iterative tomography reconstruction engine with the integration of perception knowledge. PFITRE combines the capabilities of a deep neural network in imaging correction, restoration, and inpainting with the generality and robustness of an iterative solving engine. A modified U-net architecture with Residual-in-Residual Dense Blocks incorporating dilated convolution layers has been developed to enhance the network's receptive field, thereby improving its performance. This modification enables the network to effectively improve reconstructions across different length scales. Synthetic data tests demonstrate that the modified U-net outperforms other U-net architectures across various metrics, including L1, VGG, SSIM, and PSNR.

Stability is a crucial attribute of the network, necessitated by its repeated applications in the iterative process. To ensure stability, we include ground truth-ground truth image pairs in the training dataset and implement an identity loss for its training. Transfer learning on a second training dataset addresses additional imperfections such as noise, misalignment, and intensity normalization issues, enhancing the network's performance on experimental data where the missing wedge is not the sole source of imperfection.

We employed Alternating Directions of Method of Multiplier as the iterative engine. A plug-and-play scheme is adopted to integrate the modified U-net as a regularizer into the iterative process. This approach offers simplicity in implementation by eliminating the need for an explicit regularization function. Focus on image-to-image correction, rather than learning the underlying physics mapping sinogram to image, makes our network effective in learning common features of samples, leveraging existing architecture, and pre-training. The interplay between the image and physics domains ensures perceptually excellent results with data consistency, minimizing over-correction issues. We demonstrated the convergence of the iterative process and illustrated gradual improvements with iterations.

Extensive verification on various samples, including integrated circuits, Li-ion battery electrodes, and porous copper samples, showcases the effectiveness, generality, and robustness of PFITRE. Even when the latter two samples were not part of the training dataset, PFITRE produces significantly improved reconstructions with limited-angle projections compared to other methods. Notably, PFITRE achieves satisfying reconstruction results even in the most challenging case with projections reduced to a range of ~70°. Our method also enhances reconstruction quality for sparse datasets, despite not being specifically trained for sparsity, highlighting its effectiveness in capturing relevant features and addressing similarity in the problem.

However, challenges remain. PFITRE currently operates on a 2D model treating slices in a 3D volume separately, potentially causing normalization issues along the stacking direction. Expanding the model to 3D would utilize prior knowledge along the third dimension and address slice-to-slice dependence, albeit at the cost of increased computational complexity. With increasing parameters in 3D model and a more complex architecture, generating a training dataset with sufficient diversity is of a challenge. Certain artifacts not included in the training dataset, such as ring artifacts induced by bad pixels and sample drifting, are not well eliminated. This indicates the method's limitation-it can only correct the type of artefacts that have been seen before.  Further development requires a more diverse training dataset covering a wider range of artifacts to enhance applicability. Efficient incremental learning with minimal training effort is another area of focus for investigation.

## 4. METHODS
### 4.1. Optimization problem in tomography

Tomographic reconstruction can be considered to find a solution to the following equation:

$$Rx = b \tag{1}$$

Here, $R$ denotes the projection matrix or forward projection operator, $b$ represents the acquired projection data from the detector, and x is the image pixels to be reconstructed. Image reconstruction involves reconstructing the unknown x from the projection data b and the projection matrix $R$. See Figure 1(A) for the schematic. When a sufficient number of projections with low noise are available, a direct method, such as Gridrec[9] or filtered back-projection (FBP)[32], suffixes to obtain a solution to Eq. (1). However, in the case of an imperfect dataset with noise or missing projections, iterative reconstruction algorithms typically yield more satisfactory results. These algorithms aim to solve an optimization problem:

$$\hat{x} = \underset{x}{argmin} \frac{1}{2}\|Rx - b\|_2^2 + g(x) \tag{2}$$

In this equation the first term, $\frac{1}{2}\|Rx - b\|_2^2$, quantifies the error of the solution to the measured data, b. Depending on the statistical model, a different functional form from L2 norm may be used. The second term, $g(x)$, represents a regularizer that enforces the solution to exhibit certain desired properties by introducing additional constraints. A frequently employed regularization function is total variation, which suppresses shot noise in the solution. In our approach, we aim to devise a "smart" regularizer capable of capturing all desired properties from prior knowledge, emulating human expertise. While explicit modeling of this perceptual knowledge is challenging, it can be effectively learned by DNNs. To this end, we design a modified U-net and train it with synthetic data to acquire complex perceptual knowledge. Subsequently, we integrate this knowledge into the iterative tomography reconstruction engine using an implicit regularization function.

### 4.2. Iterative reconstruction engine with a DNN

There exist numerous approaches to solving Eq. (2), leading to various iterative tomography algorithms[27,29]. Among these methods, ADMM is notably popular due to its robustness and easiness for implementation[27]. For completeness, here we give a brief derivation of the algorithm. In ADMM, the variable is split into two parts, one for each term, with the constraint of being equal,

$$\hat{x}, \hat{z} = \underset{x,z}{argmin}\{\frac{1}{2}\|Rx - b\|_2^2 + g(z)]\} \tag{3}$$
$$\text{subject to } x = z,$$

By introducing this constraint, Eq. (2) is transformed into a constrained optimization problem, solvable through the method of Lagrangian multipliers. Augmented Lagrangian adds robustness to the algorithm by introducing an additional penalty term, expressed as,

$$L_\tau(x, z) = \frac{1}{2}\|Rx - b\|_2^2 + g(z) + y^T(x - z) + \frac{\tau}{2}\|x - z\|_2^2 \tag{4}$$

Where τ > 0 is an auxiliary parameter, superscript T denotes the transpose operator, and y represents the dual variable. At each iteration, minimizations over x and z are achieved by fixing the other variables, followed by the update of the dual variable, y. This decomposition into sub-problems continues iteratively until convergence. The algorithm can be expressed as follows:

$$x_{k+1} = \underset{x}{argmin}\ \{\frac{1}{2}\|Rx - b\|_2^2 + y_k^T x + \frac{\tau}{2}\|x - z_k\|_2^2\} \tag{5}$$

$$z_{k+1} = \underset{z}{argmin}\ \{g(z) - y_k^T z + \frac{\tau}{2}\|x_{k+1} - z\|_2^2\} \tag{6}$$

$$y_{k+1} = y_k + \tau(x_{k+1} - z_{k+1}) \tag{7}$$

For convenience, we combine linear and quadratic terms and use a scaled dual variable, $u = \left(\frac{1}{\tau}\right)y$. This leads to a slightly different but equivalent form,

$$x_{k+1} = \underset{x}{argmin}\ \{\frac{1}{2}\|Rx - b\|_2^2 + \frac{\tau}{2}\|x - (z_k - u_k)\|_2^2\} \tag{8}$$

$$z_{k+1} = \underset{z}{argmin}\ \{g(z) + \frac{\tau}{2}\|z - (x_{k+1} + u_k)\|_2^2\} \tag{9}$$

$$u_{k+1} = u_k + (x_{k+1} - z_{k+1}) \tag{10}$$

In these steps, x-minimization can be solved by many standard techniques, while z-minimization poses a challenge due to the lack of an explicit functional form for g(z). Considering its effect of being a special filter based on the regularization term, we argue that there exists an implicit function, resulting in a solution for z-minimization being the output of a DNN. This concept is known as the Plug-and-Play (PnP) scheme and has demonstrated success in various image processing applications. Chan *et al.* combined the BM3D denoiser with the ADMM algorithm for denoising and image super-resolution[33]. Zhang *et al.* first incorporated a CNN-based denoiser into the ADMM algorithm[34]. Recently, Zhu *et al.* incorporated a diffusion model, with generative power, as regularization for image restoration, showcasing its superior capability in image deblurring, inpainting, and super-resolution[35].

In light of this, rather than calculating the gradient of g(z) and solving the sub-optimization problem of Eq. 10, we update z-step using,

$$z_{k+1} = DNN(x_{k+1} + u_k), \tag{11}$$

ADMM with PnP eliminates the need for defining an explicit regularizer and simplifies z-minimization. This integration allows the incorporation of a DNN into the iterative solving process with minimal effort. The DNN restores, corrects, and inpaints the corrupted reconstruction, while the x-minimization step ensures the plausibility of these changes by enforcing consistency with the physical model and the measured data. The interplay between these two processes leads to a solution satisfying both data consistency and perceptual expectations (see Figure 1(c)). From a statistical standpoint, we maximize the posterior probability and leverage strong priori to select the most probable solution, where the prior distribution is encoded in the DNN.

### 4.3. Design of the network architecture

The network architecture design is illustrated in Figure 1(c). To capture features at different scales, we adopt a typical image translation network architecture: U-net convolutional neural network (CNN) that follows an encoder-decoder scheme. Built upon the U-net, we incorporate four Residuals in Residual Dense Blocks (RRDB)[36] at the bottleneck to manage latent space information. This design enables dense connections between layers and supports reusing feature maps from earlier layers efficiently. It is inspired by previous work in CycleGAN[37] and UVCGAN[38], which efficiently enhance non-local pattern learning. Among four RRDB blocks, two consist of regular 3×3 convolution layers, and the other two comprise dilated convolution with dilation rates of 2, 4, and 8 in each layer. The design of RRDB with dilated convolution aims to efficiently expand the receptive field and fuse multi-scale information, particularly beneficial for line-shaped features in IC samples. Additionally, we replaced the max-pooling layers with 2×2 convolution layers with a stride of 2, following previous denoising tasks and preventing checkerboard artifacts[39,40]. To enhance convergence properties and prevent overfitting, we utilized a pre-trained VGG network[41,42] to initialize the weights of the encoder in our modified U-net architecture. While VGG employs the ReLU activation function, we replaced it with GELU to avoid issues related to 'dead neurons' and zero gradients.

### 4.4. Loss function

Our goal is to remove artifacts, inpaint missing details, and restore distortions of the reconstruction, meanwhile maintaining the data consistency in an iterative manner. To achieve this we employed a supervised learning method, with training dataset containing image pairs of ground truth and corrupted reconstruction. The loss function for network training is defined as:

$$L = \lambda_{L1} \times L_1 + \lambda_{VGG} \times L_{VGG} + \lambda_{identity} \times L_{identity}. \qquad (12)$$

The $L_1$ norm and VGG loss[41] were used to measure the mean absolute difference and feature difference between network output and the ground truth respectively, and are widely adopted metrics for supervised learning in various image tasks. The identity loss quantified the disparity between the 1st-time and the 2nd-time outputs from the network. In this process, the 1st-time output was fed back into the network, and its subsequent network output was designated as the 2nd-time output. Ensuring consistency between the two-time outputs is crucial when integrating the network into an iterative correction process to minimize overcorrection and maintain stability.

To quantitatively compare the performance of the modified U-net with other networks, we utilized various metrics, including $L_1$, VGG, Peak Signal-to-Noise Ratio (PSNR), and the Structural Similarity Index (SSIM), to gauge the dissimilarity between the corrected image and the ground truth. PSNR measures the ratio of the maximum signal power to the noise affecting the signal, where higher PSNR values indicate better image quality. SSIM quantifies high-level features for human perception, assessing visual appeal even in the presence of imperfect images. It falls within the range of [0, 1], with 0 signifying no similarity and 1 denoting that two images are identical. VGG metrics compare feature differences, making it robust for noisy images and artifacts compared to SSIM. Smaller VGG values indicate better image quality.

The entire workflow was implemented using Python, and the neural network was built using the PyTorch framework[43]. Training was carried out on a single Nvidia Tesla V100 graphics card. Initially, we conducted 50 epochs with a fixed learning rate of 2e-4 on the dataset containing only missing-angle artifacts. Subsequently, the network underwent fine-tuning for 10 epochs using a dataset containing various types of artifacts. To expedite the learning process and reduce GPU memory requirements, we implemented the automatic mixed-precision method during training.

Due to the use of GPU and its rapid convergence properties, our iterative method requires a similar processing time as other iterative reconstruction algorithm. The reconstruction comparison was conducted on a GeForce RTX 3060, with the test image being approximately 150 × 150 pixels. The GANrec method takes 61 seconds to reconstruct one image after 5000 iterations. The running time for MLEM, FISTA, and our method is less than a second for processing each image after 50, 50, and 15 iterations, respectively. It can be faster if a physical model-based linear solver (x-minimization) is also implemented on the GPU. Currently, only network-based regularization (z-minimization) runs on the GPU.

### 4.5. Training and test datasets

The modified U-net is trained using a supervised learning method with image pairs consisting of ground truth and reconstructions with artifacts. The latter is generated using the generalized minimal residual iteration algorithm that is available in SciPy[28]. Projections with angles spanning from 90° to 160° with a 1° angular increment are used to generate the synthetic data. Given the iterative nature of the network's application, mitigating over-correction and maintaining stability are paramount. Therefore, image pairs of ground truth to ground truth are included in the training data. The training dataset encompasses high-resolution natural images collected from 256 object categories[44], texture images[45], microscopic images of integrated circuits (ICs)[46], and simulated semiconductor patterns. Data augmentation techniques, such as small-angle random rotation, horizontal and vertical flipping, and cropping to different sizes, are applied. The focus on ICs stems from the strong demand for chip imaging using limited-angle x-ray nano-tomography[8,47].

All images are resized to 320×320 pixels, and their intensities are normalized individually to a consistent scale based on each image's mean and standard deviation. In total, the training dataset comprises approximately ~33000 image pairs containing limited-angle artifacts, while the test dataset consists of around 6500 image pairs. No images from the test dataset were included in the training process. The network architecture comparison is based on this synthetic test dataset.

To address the impact of other types of imperfections encountered in an experiment, we generated another synthetic image dataset. We introduced Poisson noise, Gaussian noise, a combination of both, misalignment, and varying intensity to sinograms generated from forward projections of the ground truth. The Gaussian noise was generated with a mean of 0, and its standard deviation (sigma) was randomly selected between 0 and 1 for each synthetic sinogram created. Subsequently, we reconstructed images from these synthetic sinograms and utilized them as a new synthetic dataset for transfer learning. In total, we created approximately 30,000 image pairs containing a variety of artifacts.

### 4.6. Experimental data collection

To validate the effectiveness of PFITRE, we tested it on various experimental datasets with different sample types, modalities and acquisition modes. They include x-ray fluorescence (XRF) imaging of IC samples taken at the Hard X-ray Nanoprobe (HXN) beamline, and absorption-based imaging of porous

Cu and $LiMn_2O_4$ electrodes taken at Full Field X-ray Imaging (FXI) beamline at the National Synchrotron Light Source II (NSLS-II) of Brookhaven National Laboratory. HXN utilizes multilayer Laue lenses to achieve an imaging resolution of about 10 nm in scanning mode[48], while FXI employs a Fresnel zone plate to do fast snapshot imaging with a resolution of about 30 nm[49]. See Figure 1(b) for the schematics of the two imaging modes. Cylindrical and planar samples for tomography measurements were prepared using a focused ion beam with scanning electron microscopy (FIB-SEM, Helios dual beam, FEI).

For the two cylindrical IC samples, projection images were taken over 180 degrees, and were used to produce a reference reconstruction as the ground truth. Images were acquired at every one degree for sample A, and every two degrees for sample B. Grid scans of 160 x 100 were performed at 12 keV for A and 167 x 94 for B, by collecting XRF signals and transmission signals from the sample The step size is 15 nm in both cases, resulting in a scanned area of 2.4 × 1.5 µm² and 2.5 × 1.4 µm², respectively. We use various subsets of full ones to emulate the effects of limited-angle tomography and validate the effectiveness of PFITRE.

Due to the geometrical constraint, projection XRF images for the planar IC sample were acquired over 130 degrees at every one degree. Each frame covers an area of 6 × 1.02 µm² with a pixel size of 15 nm. Subsets of the data with further reduced angular range were used to explore the limit of the algorithm.

For full-field tomography datasets acquired at FXI, they have a field of view of 41.6 µm × 35.1 µm (horizontal × vertical) with a pixel size of 20 nm. The TXM images were collected at X-ray energies above the Mn K-edge (6.539 keV) to maximize the contrast for $LiMn_2O_4$ materials and above the Cu K-edge (8.979 keV) to maximize the contrast for Cu materials.

In each dataset, images were aligned by the StackReg[50] plugin in the freeware Imagej[51]. The XRF fitting was conducted using PyXRF package[52].

ACKNOWLEDGEMENTS

This research used Hard X-ray Nanoprobe (HXN) Beamline (3-ID), Full-Field X-ray Imaging (FXI) beamline (18-ID) of the National Synchrotron Light Source II, a US Department of Energy (DOE) Office of Science User Facility operated for the DOE Office of Science by Brookhaven National Laboratory under contract no. DE-SC0012704. This work is supported partly by BNL LDRD funding (24-067). We thank the user support provided by the CFN staff Kim Kisslinger and Fernando Camino for FIB-SEM training and help with sample preparation. We acknowledge the sample preparation of porous Cu and $LiMn_2O_4$ battery electrodes by Qingkun Meng and Cheng-Hung Lin from the Chen-Wiegart group at Stony Brook University and Brookhaven National Laboratory. We also thank HXN beamline staff Ajith Pattammattel and Xiaojing Huang for their support of XRF tomography measurement setup and data analysis.

AUTHOR CONTRIBUTIONS

H. Y conceived the original idea and supervised the entire project. H. Y and Y. S. C secured resources and funding. C. Z created the training datasets and designed the network, with contributions from H. Y, M. G, and X. Y. C. Z conducted the network training. C. Z and H. Y performed the experiments, collected the data, and did data analysis. All authors contributed to the manuscript preparation.

COMPETING INTERESTS

The Authors declare no Competing Financial or Non-Financial Interests.

CODE AVAILABILITY

Network architecture, example data, and demonstration code can be found at https://github.com/chonghangzhao/PFITRE.

DATA AVAILABILITY

Model can be downloaded at https://drive.google.com/file/d/1rqop4dAZ5QSjZluPkQnnMj5Qkmn5gtKo/view?usp=drive_link. Training datasets and code are available from the corresponding author upon reasonable request.

Supplementary information

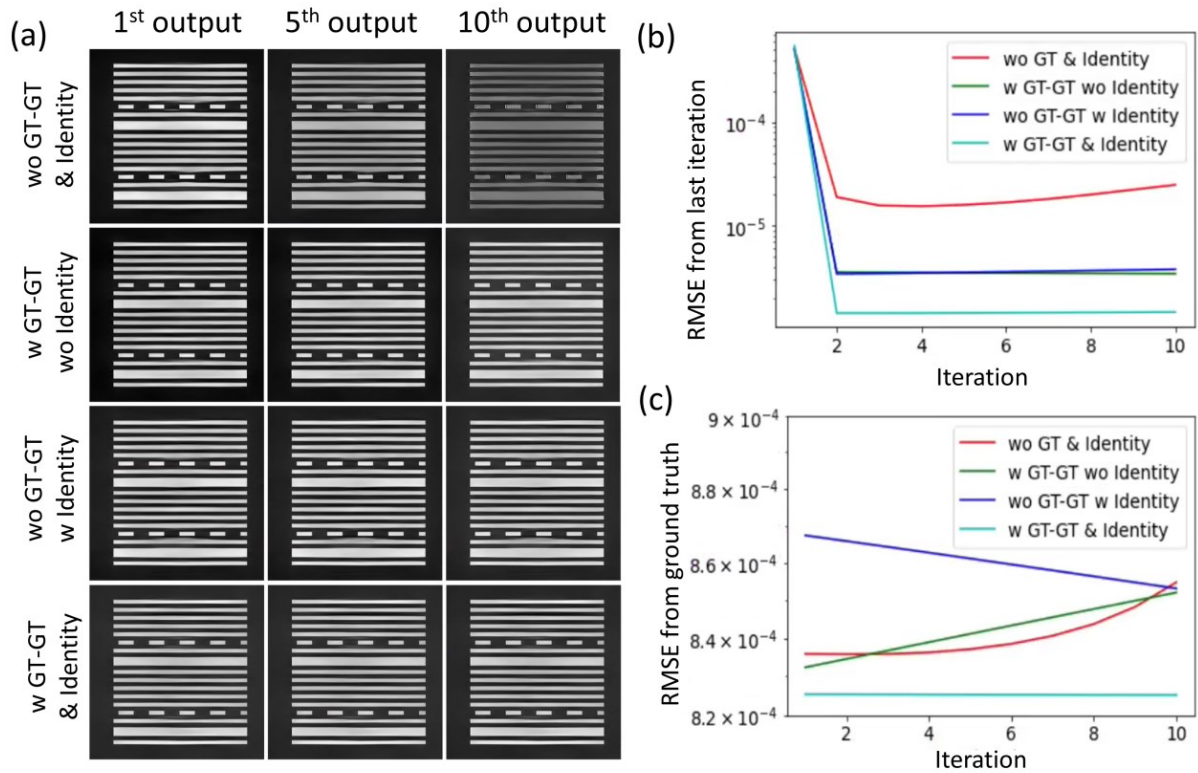

Figure S1. Iterative post-processing correction on synthetic images with limited-angle artifacts by network trained with different datasets and loss functions. The baseline is training with L1 and VGG losses, without ground truth-ground truth (GT-GT) image pairs. (a) Iteratively feed output back into the network after 1, 5, and 10 times. The involved comparisons included a network trained without GT-GT pairs, and without identity loss as the baseline; a network trained with GT-GT pairs but no identity loss, a network trained without GT-GT pairs but including identity loss, a network trained with GT-GT pairs and identity loss. (b) RMSE of each output from the last iteration. (c) RMSE of each iterative output from the ground truth.

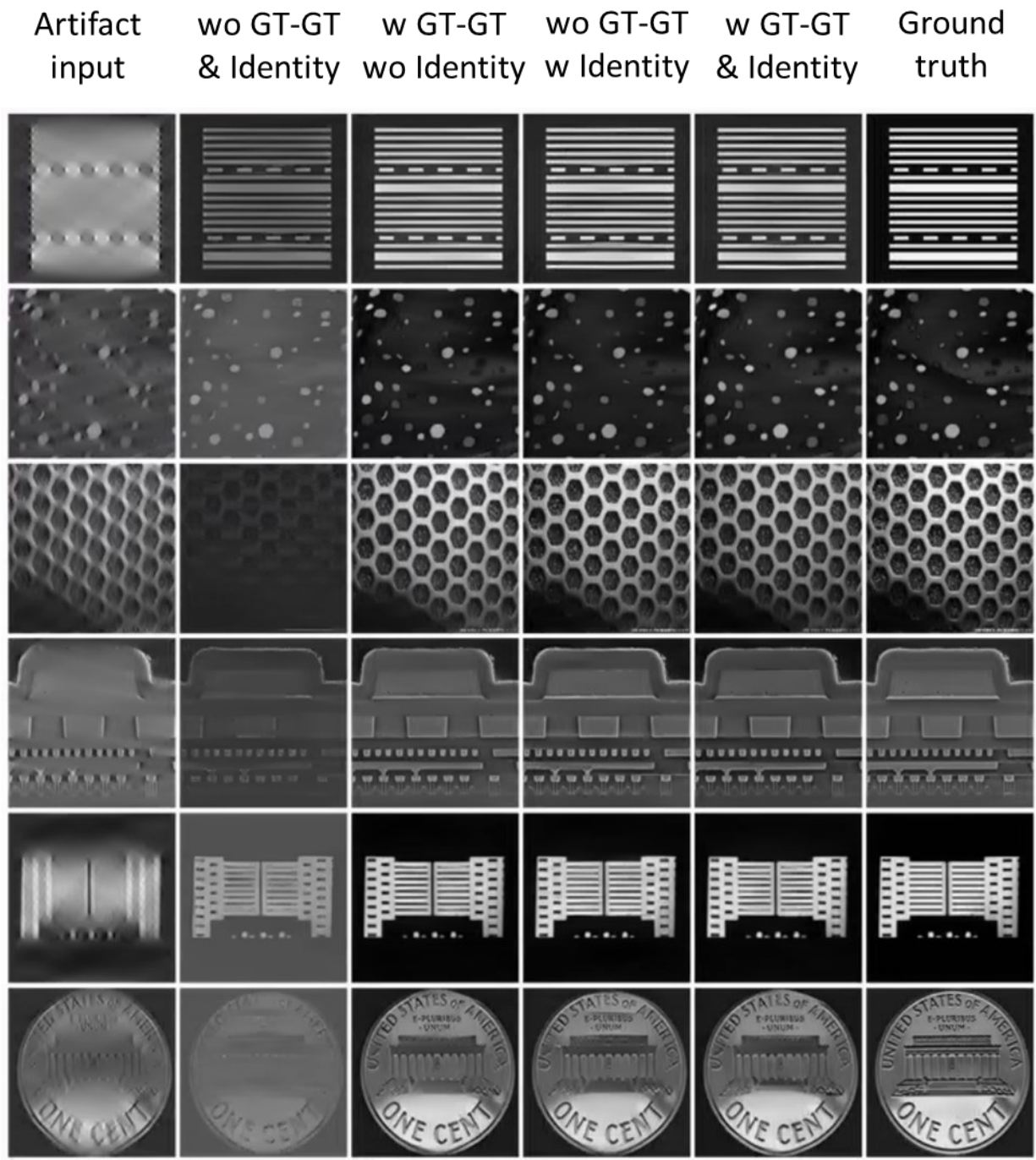

Figure S2. Iterative post-processing correction was performed ten times on synthetic images with limited-angle artifacts using networks trained with different datasets and loss functions. The baseline was training with $L_1$ and VGG losses, without ground truth-ground truth (GT-GT) image pair. The comparisons included the following scenarios: network trained without GT-GT pair and identity loss (baseline); network trained with GT-GT pair but no identity loss; network trained without GT-GT pair but including identity loss; and network trained with GT-GT pair and identity loss.

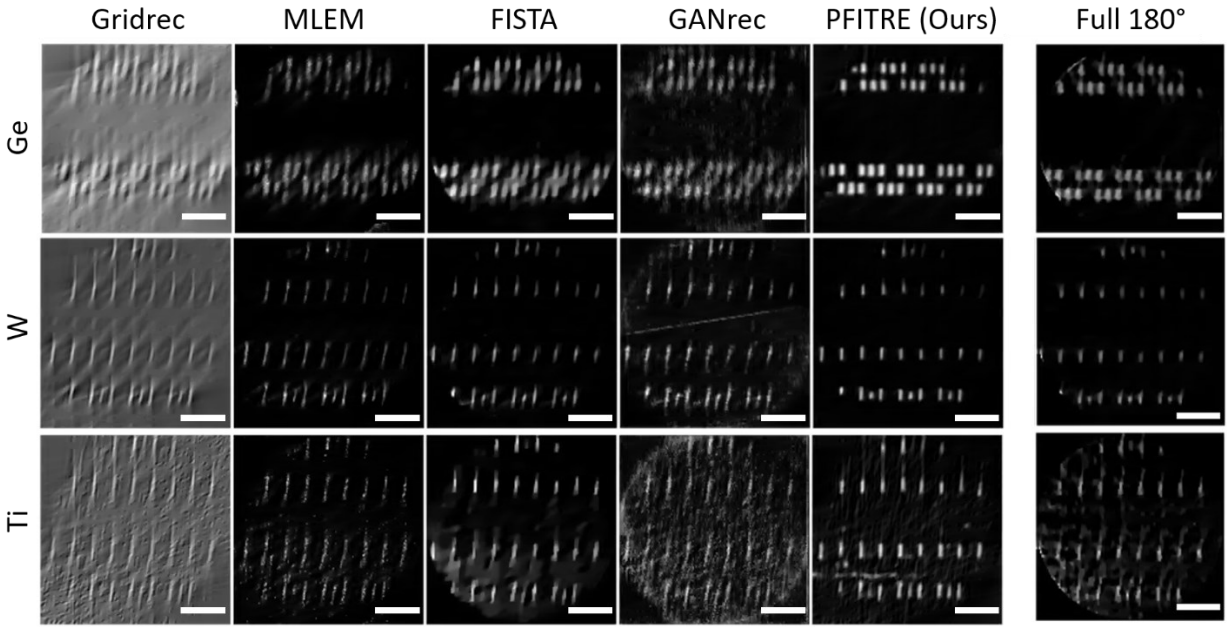

Figure S3. Reconstructed with the subset of projections using different reconstruction methods on the Ge, W, and Ti fluorescence datasets. Comparison methods include reconstructions with Gridrec, MLEM, FISTA, GANrec method, our method, and the full dataset covered 180° reconstructed by the FISTA method. All scale bars indicate 500 nm.

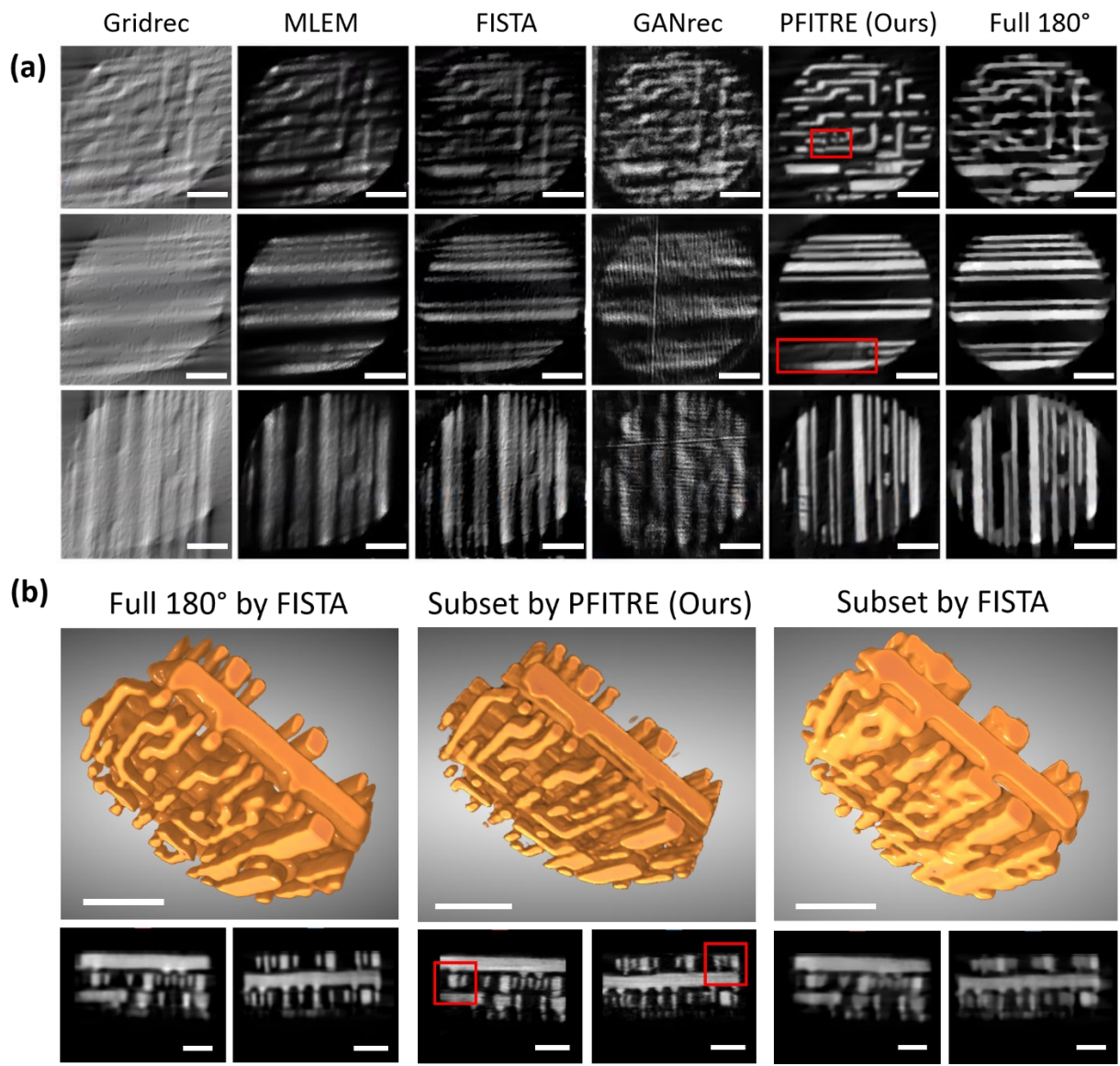

Figure S4. Reconstructed with a subset of projections using different methods on sample 'b'. (a) A comparison of reconstructions on three layers with Gridrec, MLEM, FISTA, GANrec method, our method, and the full dataset covered 180° reconstructed by the FISTA method. All scale bars indicate 500 nm. (b) 3D volume rendering of reconstruction from the 180 angular range dataset, from subset by our method, and the FISTA method.

The effect of different algorithms generated image as the initial guess

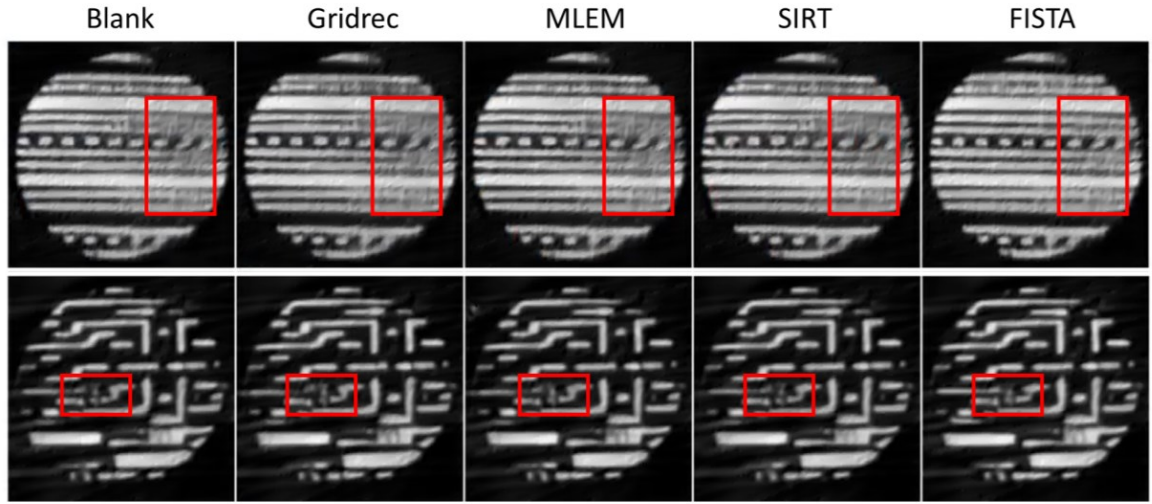

Figure S5. The effect of initial guess on our iterative reconstruction result. Reconstruction with different initial guesses from two layers of IC samples. From left to right: iterative reconstruction result with no initial guess, with initial guess reconstructed by Gridrec, with initial guess reconstructed by MLEM, with initial guess reconstructed by SIRT, and initial guess reconstructed with FISTA.

In Figure 1(c) working flow of ADMM, we observe that ADMM starts with an initial guess for iterative correction. The quality of correction is closely related to the quality of the input image into the network and can be further improved by using other advanced reconstruction results as initial guesses in our iterative method. The impact of different initial guesses on our reconstruction results is shown in *Figure S5*. It is evident that image quality slightly improves when reconstruction starts from the FISTA method. This improvement can be attributed to the provision of a better initial guess, allowing our iterative method to begin from a local minimum and iteratively approach the global minimum during optimization. In the future, other advanced one-time correction results can be incorporated as initial guesses into our iterative method for further enhancements to ensure no further artifacts are introduced.

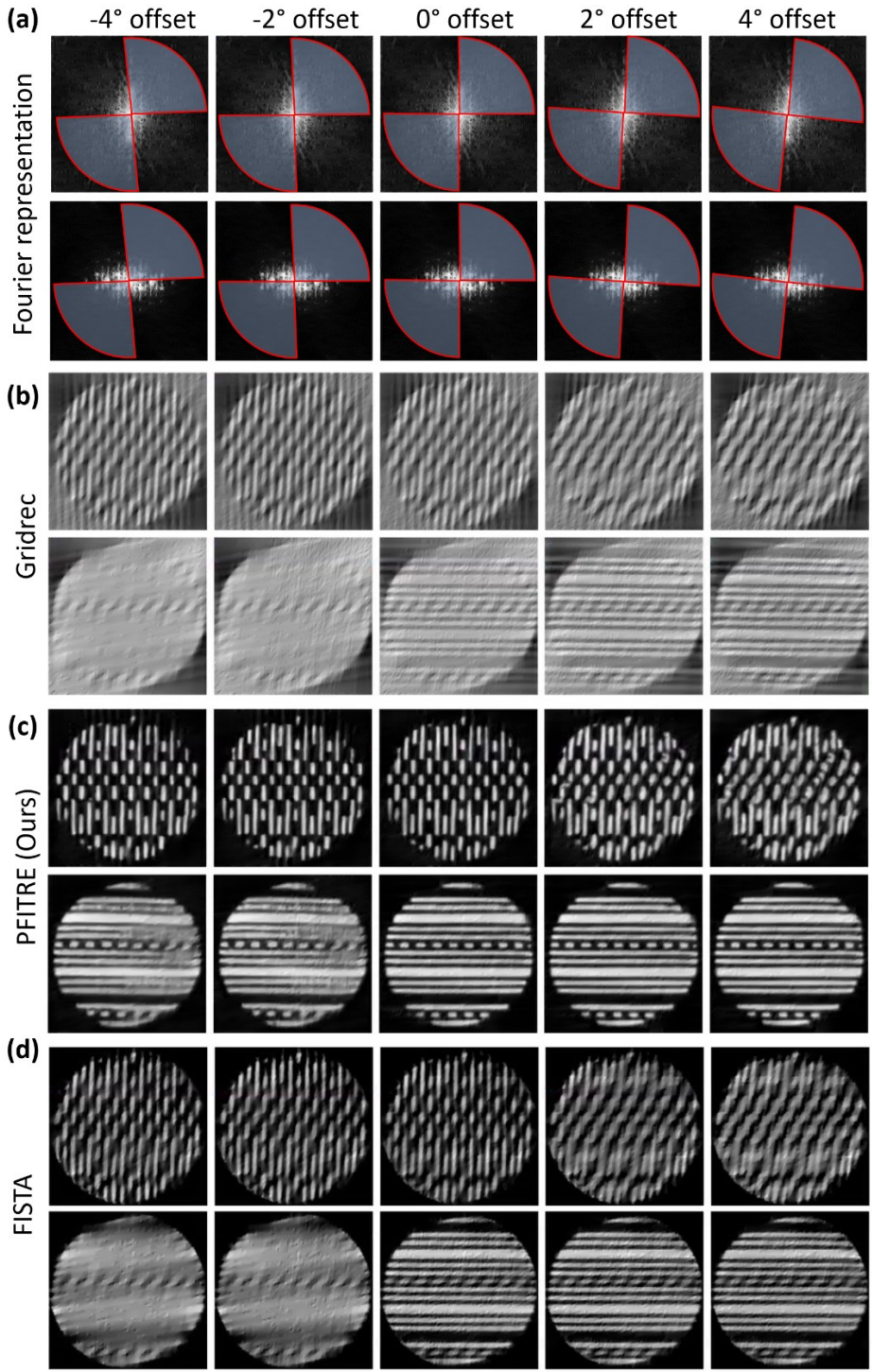

Figure S6. The effect of missing angular range concerning line feature distribution. (a) The illustration of Fourier representation of different missing angular ranges to line features. (b) the reconstruction by the Gridrec method, (c) the reconstruction by our method, and (d) the reconstruction by the FISTA method.

When measuring IC samples with line-shaped features, we must consider not only the covered angular range but also the relative converging angle with respect to features. This is because, in IC samples with line features, their information is predominantly orthogonally distributed in Fourier space. We controlled the missing angle range and adjusted the relative coverage with the orthogonally distributed information. The reconstruction by the Gridrec method, along with its corresponding Fourier space representation, is shown in *Figure S6*(a). It can be observed that a 2-degree gap at the crucial angle leads to significant distortion in the reconstructed images. Since a substantial amount of information is missing, it can hardly be addressed by conventional TV-based regularization. A comparison of TV-based iterative methods with our machine learning-based method is presented in *Figure S6* (b) and (c).

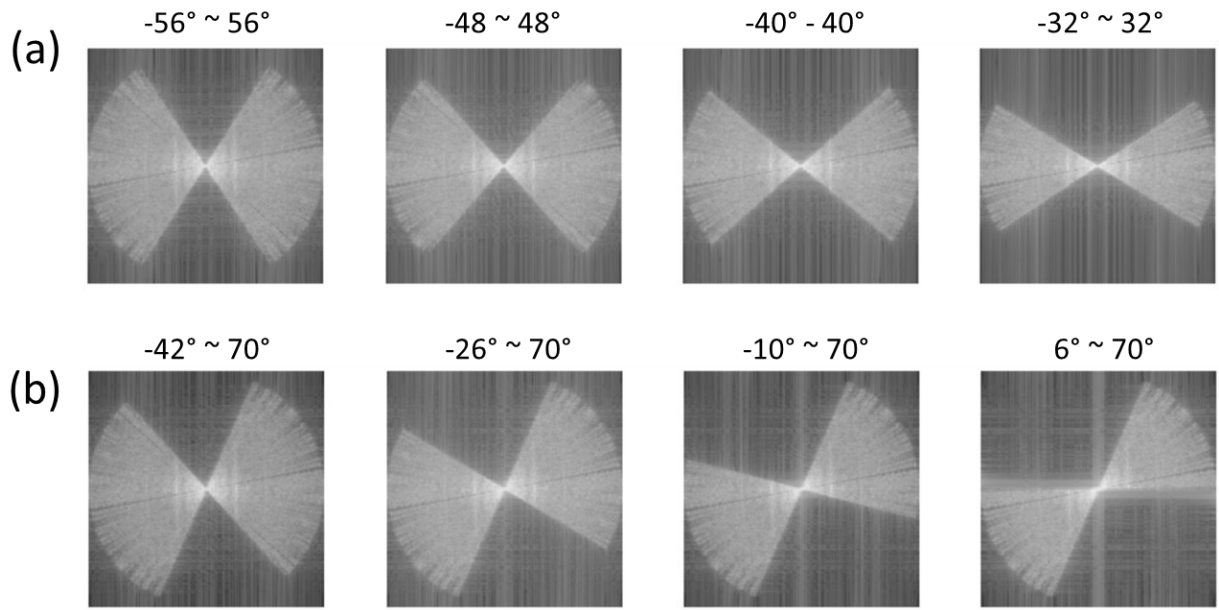

Figure S7. The illustration of Fourier representation of different angular distribution for projections. (a) Fourier representation of projections which is symmetrically distributed around 0° at different angular ranges, with 0° defined as the angle perpendicular to the length of the IC sample. (b) Fourier representation of asymmetrically distributed projections, covering angles ranging from 64° to 112°. The collected angle ranges from -58° to 70°, covering 128° in total.

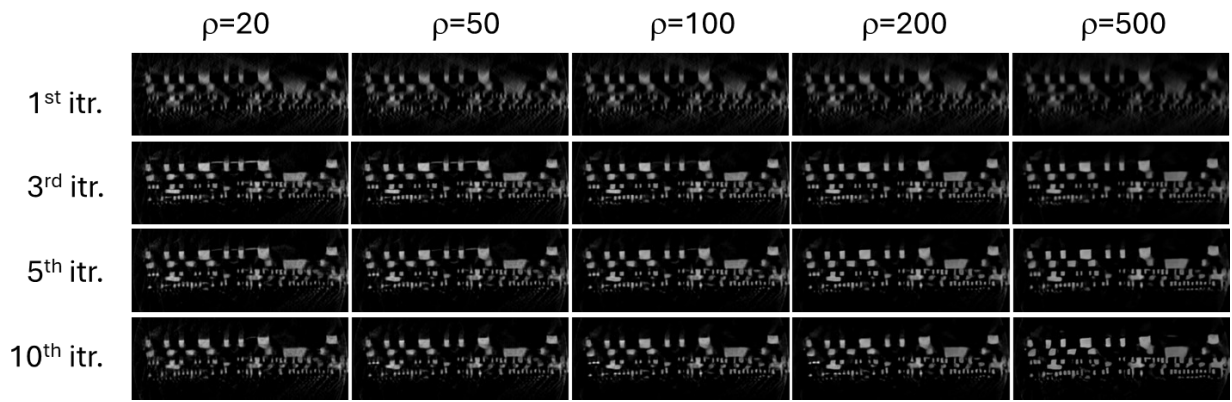

Figure S8. Impact of the weight. Row shows the reconstruction outcomes at the 1st, 3rd, 5th, and 10th iteration, and columns correspond to $\rho$ values ranging from 20 to 500. Projections were limited to $\pm 56°$. All images have the same display range.

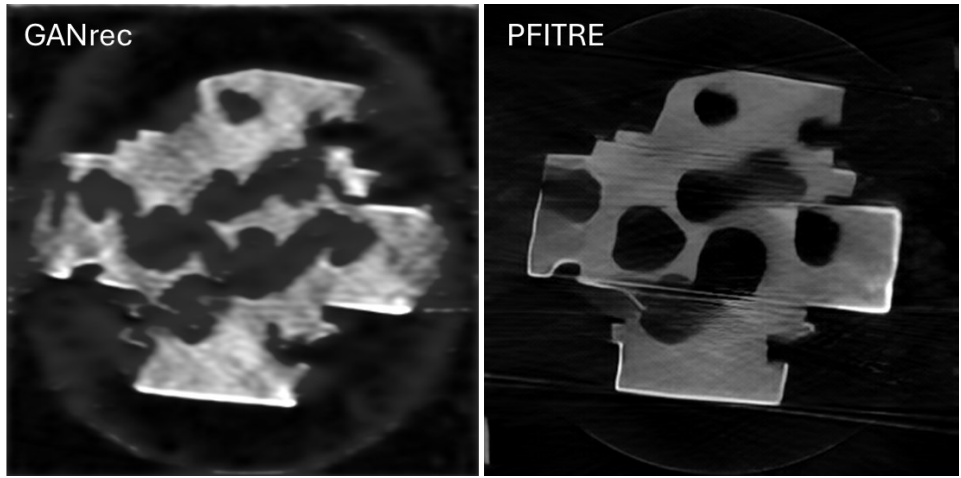

Figure S9. Comparison of reconstructions using GANrec and PFITRE on the dataset that was previously presented in Fig. 6 of reference 18. Display range is the same for both images.

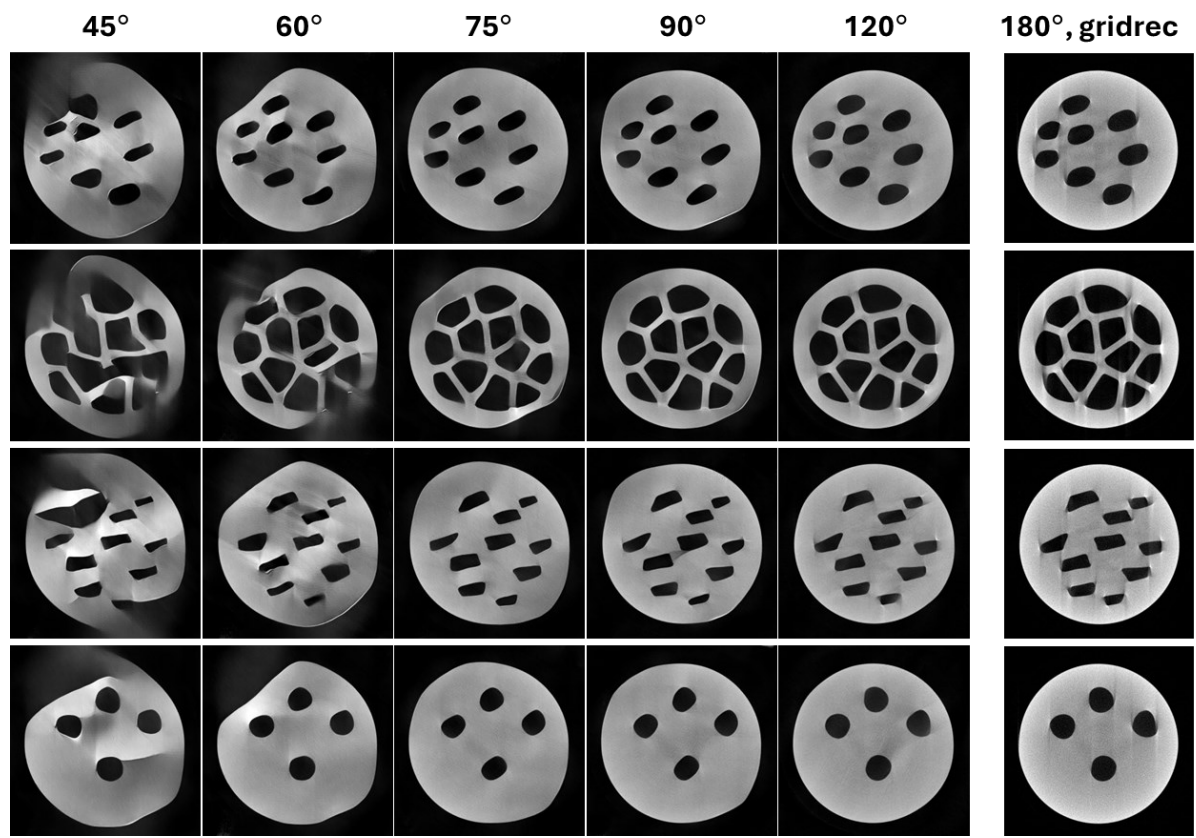

Figure S10. PFITRE reconstructions of 4 test objects from Helsinki Tomography Challenge 2022[1]. Projections cover an angular range of 45°, 60°, 75°, 90° and 120° with 0.5° increment, respectively. For a comparison, the rightmost column shows the reconstructions from projections over 180° using direct method gridrec. Display range is kept the same for all images. Note that our method was developed for parallel-beam tomography, while the data was taken in cone-beam geometry. Also note we did not retrain our network with additional training datasets provided in the contest. Therefore, these objects were unseen by the network.

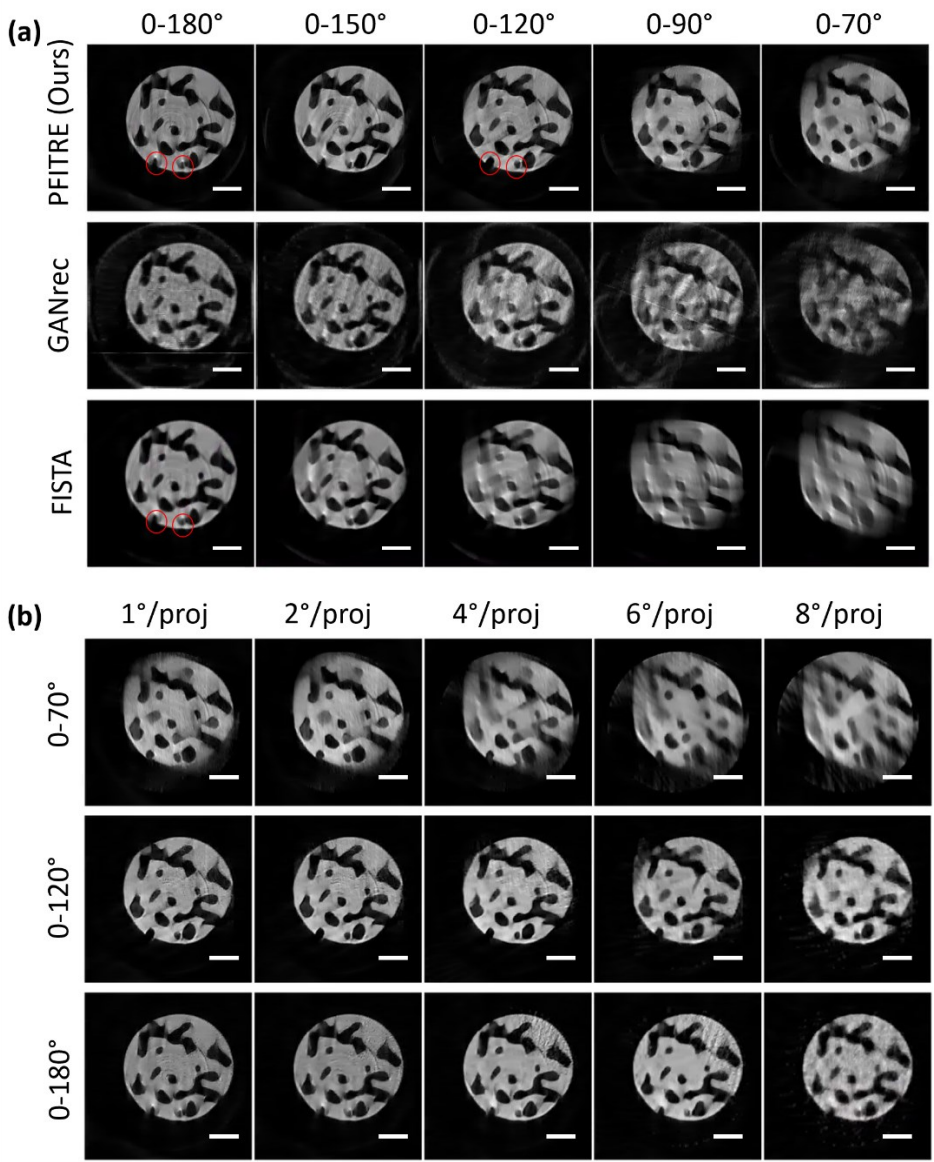

Figure S11. Comparison of reconstruction of porous Cu collected by transmission X-ray microscopy, by our iterative method, GANrec, and FISTA. (a) A comparison of reconstructions with projections covering an angular range from 70° to 180° using different methods. (b) Evaluation of performance on sparse datasets with varying angular coverage and sparsity levels. All scale bars represent a length of 5 µm.

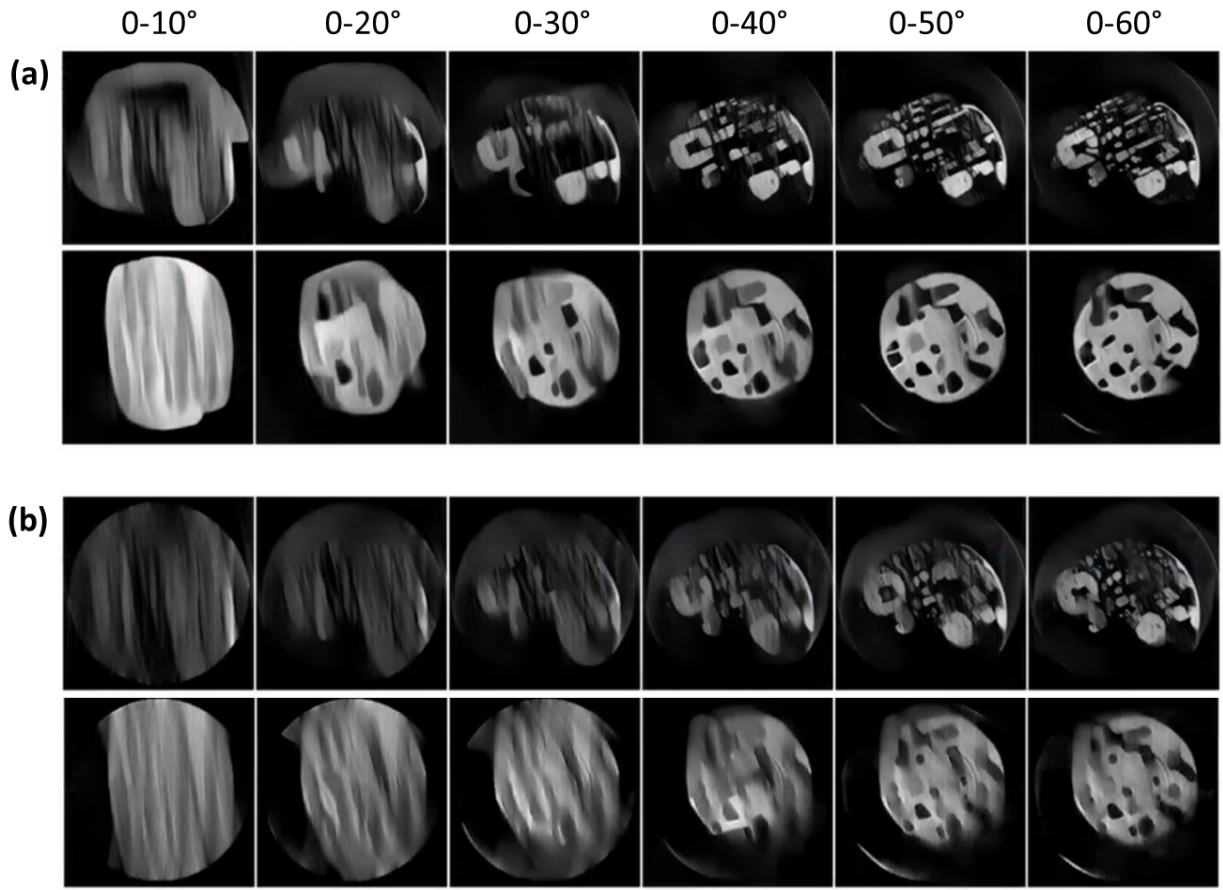

Figure S12. Reconstruction of LiMn$_2$O$_4$ electrode and porous Cu with projections covering 10 to 60° angles by networks trained with different datasets. (a) the network trained with missing angles in the range of 20°-130°. (b) the network trained with missing angles in the range of 20°-90°